\documentclass{aa}

\usepackage{graphicx,natbib}
\usepackage{bm}

\usepackage{color}
\usepackage{ulem}

\begin{document}
   \title{Are chains of type I radio bursts generated by similar processes as drifting pulsation structures observed during solar flares?}

   \authorrunning{}
   \titlerunning{Chains of type I radio bursts ...}

   \author{M. Karlick\'y}
   \offprints{M. Karlick\'y, \email{karlicky@asu.cas.cz}}

   \institute{Astronomical Institute of the Czech Academy of Sciences, Fri\v{c}ova 258, CZ -- 251 65 Ond\v rejov, Czech Republic}
   \date{Received ; accepted }


  \abstract
   {}
   {Owing to similarities of chains of type I radio bursts and drifting pulsation structures a question arises
   if both these radio bursts are generated by similar processes.}
   {Characteristics and parameters of both these radio bursts are compared.}
   {We present examples of the both types of bursts and show
   their similarities and differences.
   Then for chains of type I bursts a similar model as for drifting pulsation
   structures (DPSs) is proposed.
   We show that similarly as in the DPS model, the chains of type I bursts can be generated by the fragmented magnetic reconnection
   associated with plasmoids interactions. To support this new model of chains of type I bursts,
   we present an effect of merging
   of two plasmoids to one larger plasmoid on the radio spectrum of DPS. This process can also explain the "wavy" appearance of some
   chains of type I bursts. Then we show that the chains of type I
   bursts with the "wavy" appearance can be used for estimation of the magnetic field strength in their sources. We think that
   differences of chains of type I bursts and DPSs are mainly owing to different regimes
   of the magnetic field reconnection.
   While in the case of chains of type I bursts the magnetic reconnection and plasmoid
   interactions are in the quasi-separatrix layer of the active region in more or less quasi-saturated regime,
   in the case of DPSs, observed in the impulsive phase of eruptive flares,
   the magnetic reconnection and plasmoids interactions are in the current sheet formed under the flare magnetic
   rope, which moves upwards and forces this magnetic reconnection.}
   {}


   \maketitle

\begin{figure*}
\centering
\includegraphics[scale=0.5]{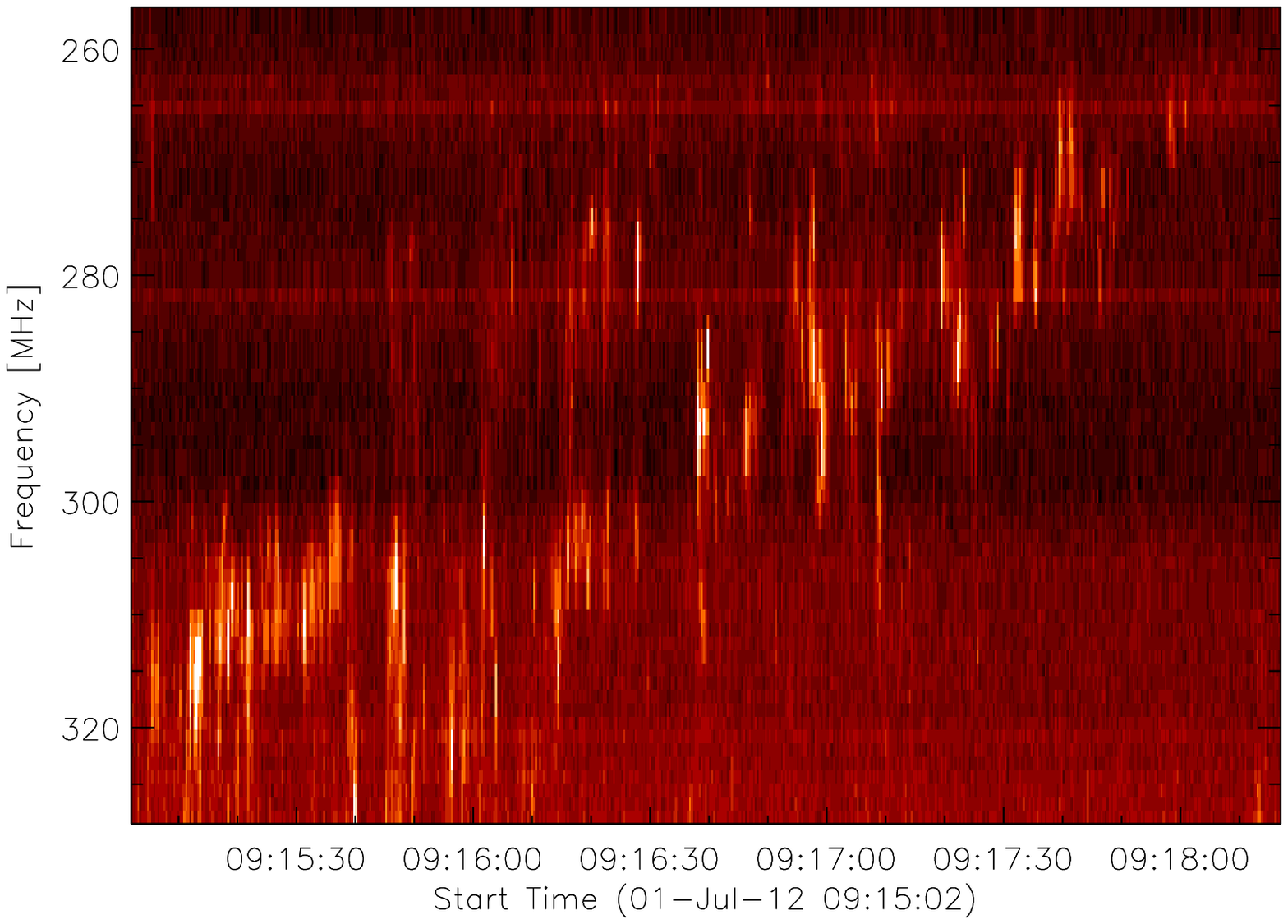}
\includegraphics[scale=0.5]{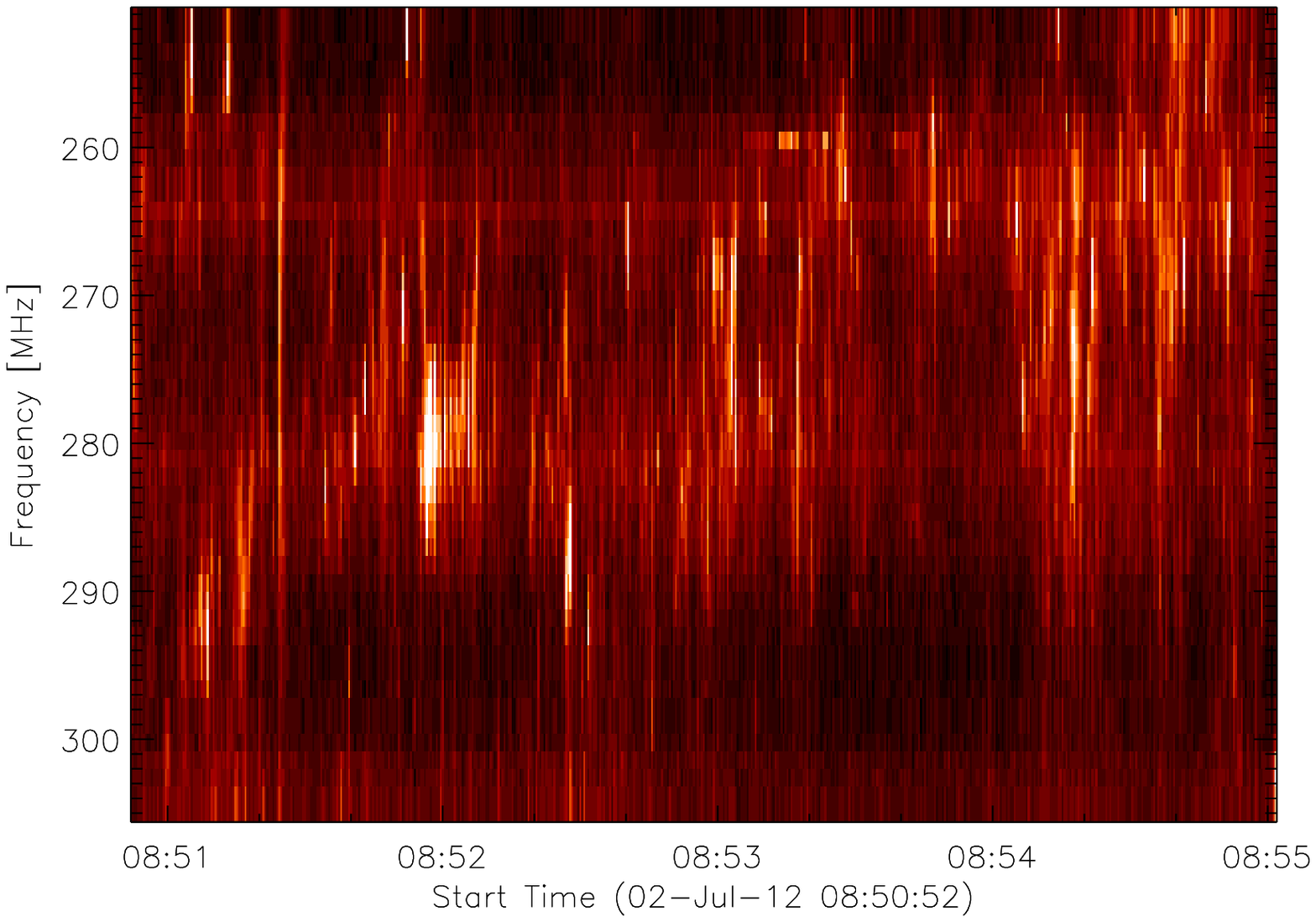}
\includegraphics[scale=0.5]{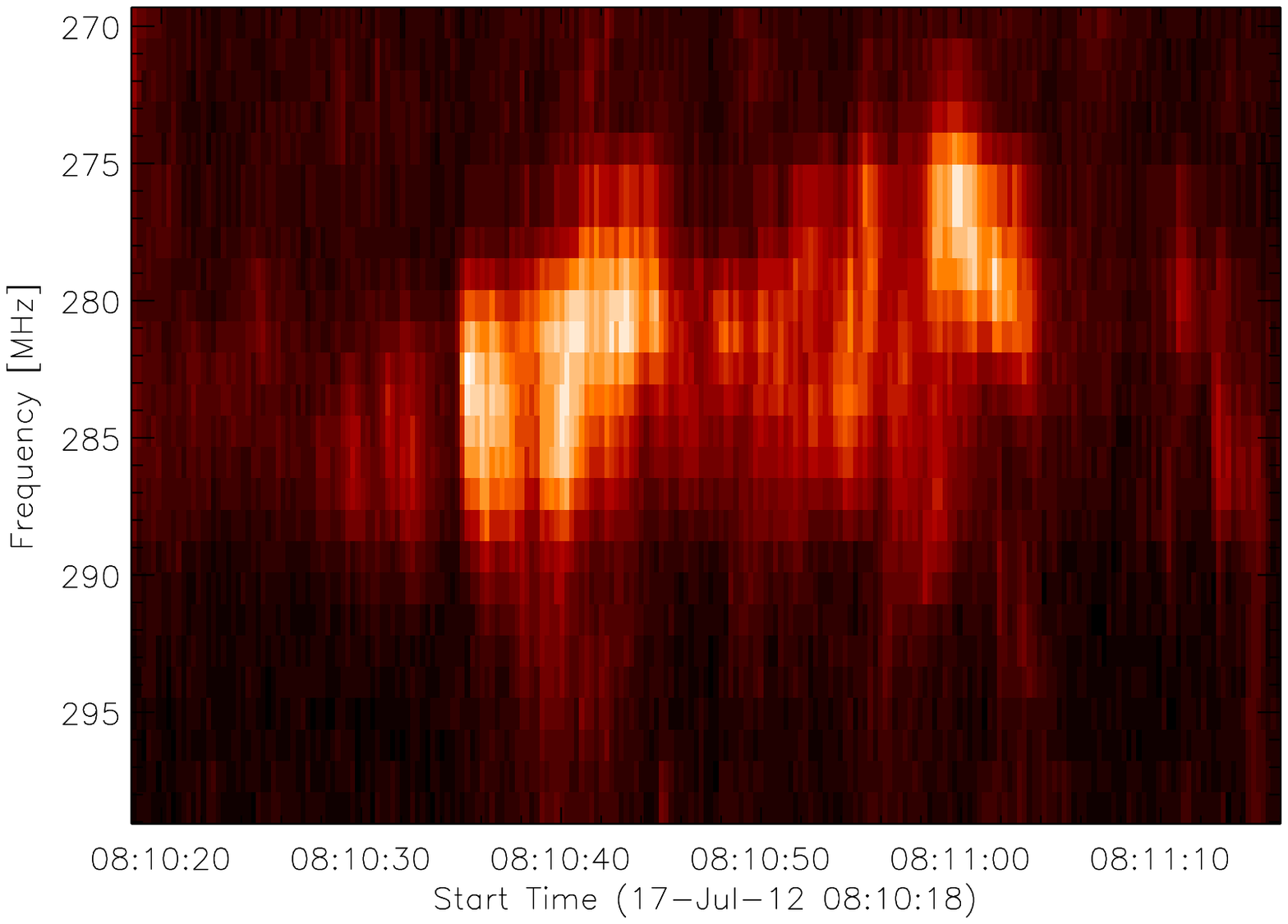}
\includegraphics[scale=0.5]{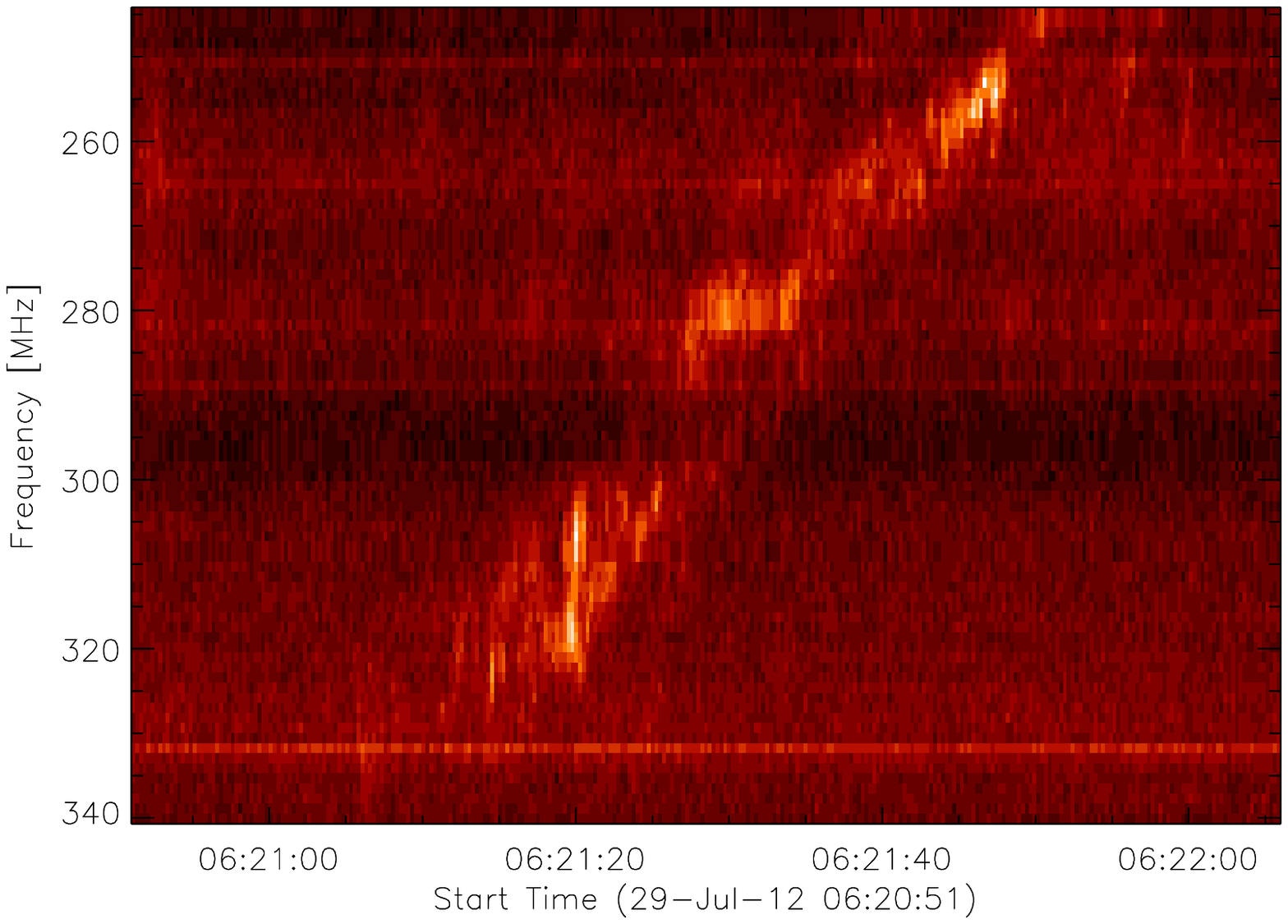}
\caption{Examples of the chains of type I radio bursts observed in the metric
range by the Callisto Trieste radiospectrograph.} \label{Fig1}
\end{figure*}

\begin{figure*}
\centering
\includegraphics[scale=0.40]{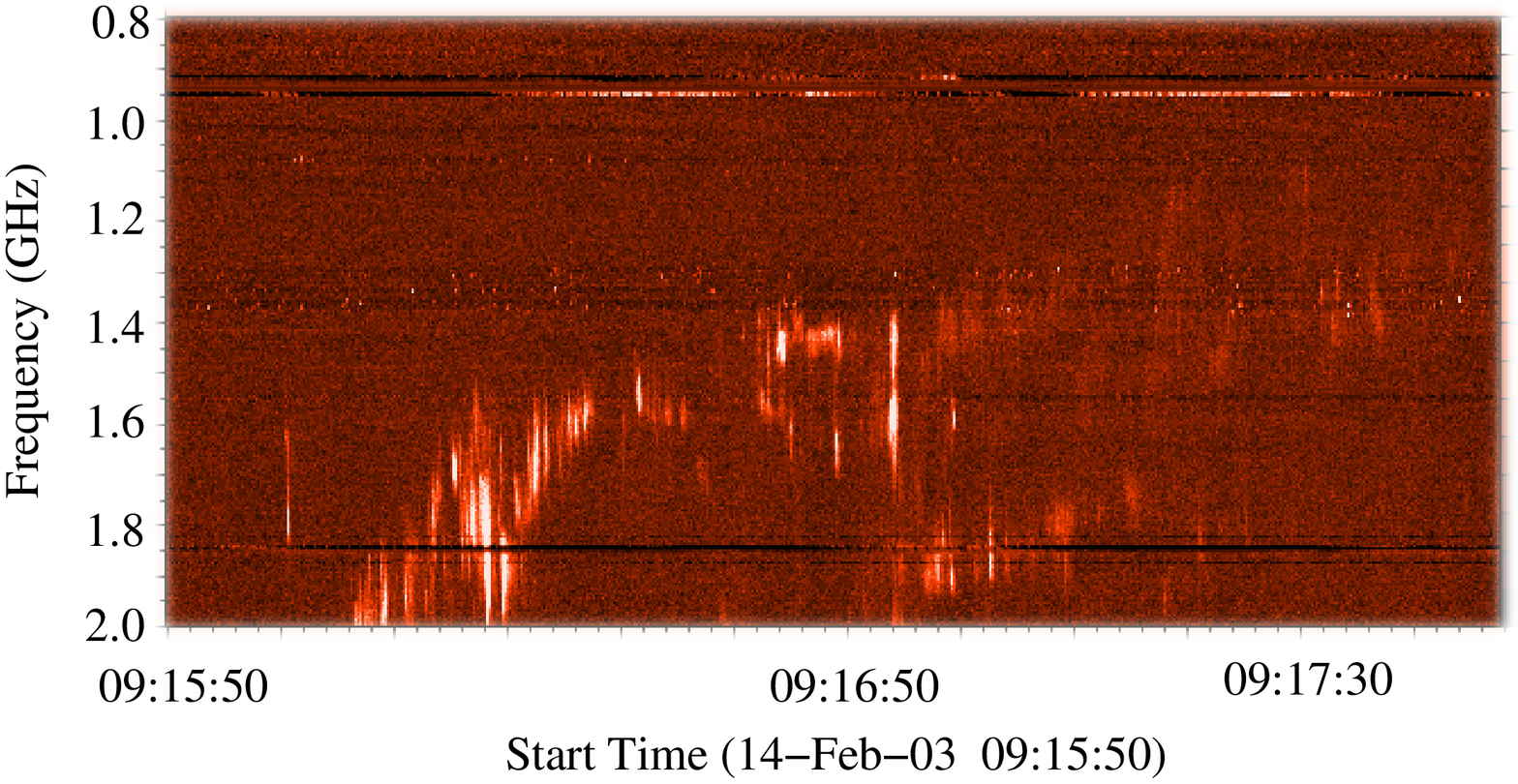}
\includegraphics[scale=0.40]{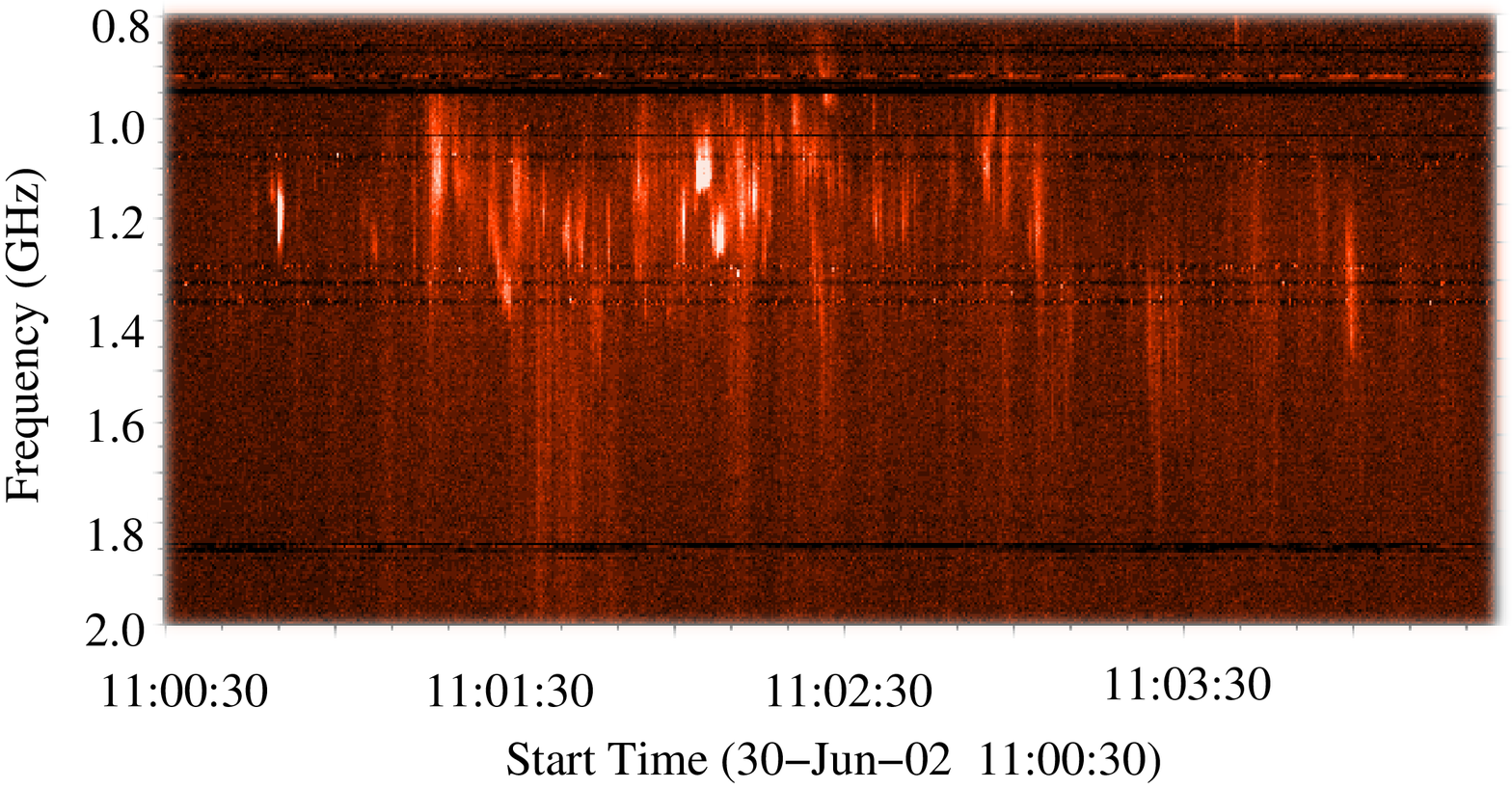}
\includegraphics[scale=0.40]{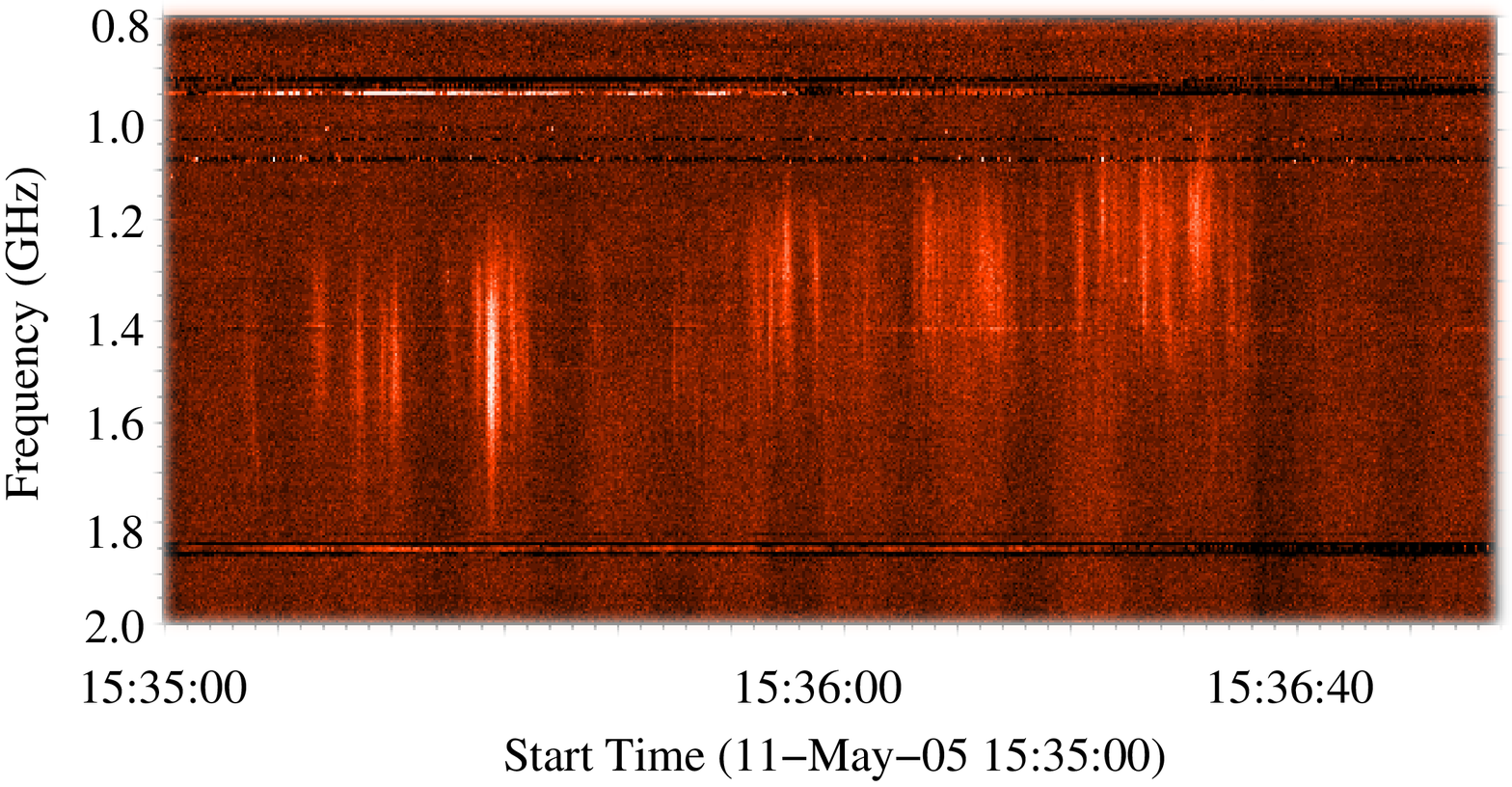}
\includegraphics[scale=0.40]{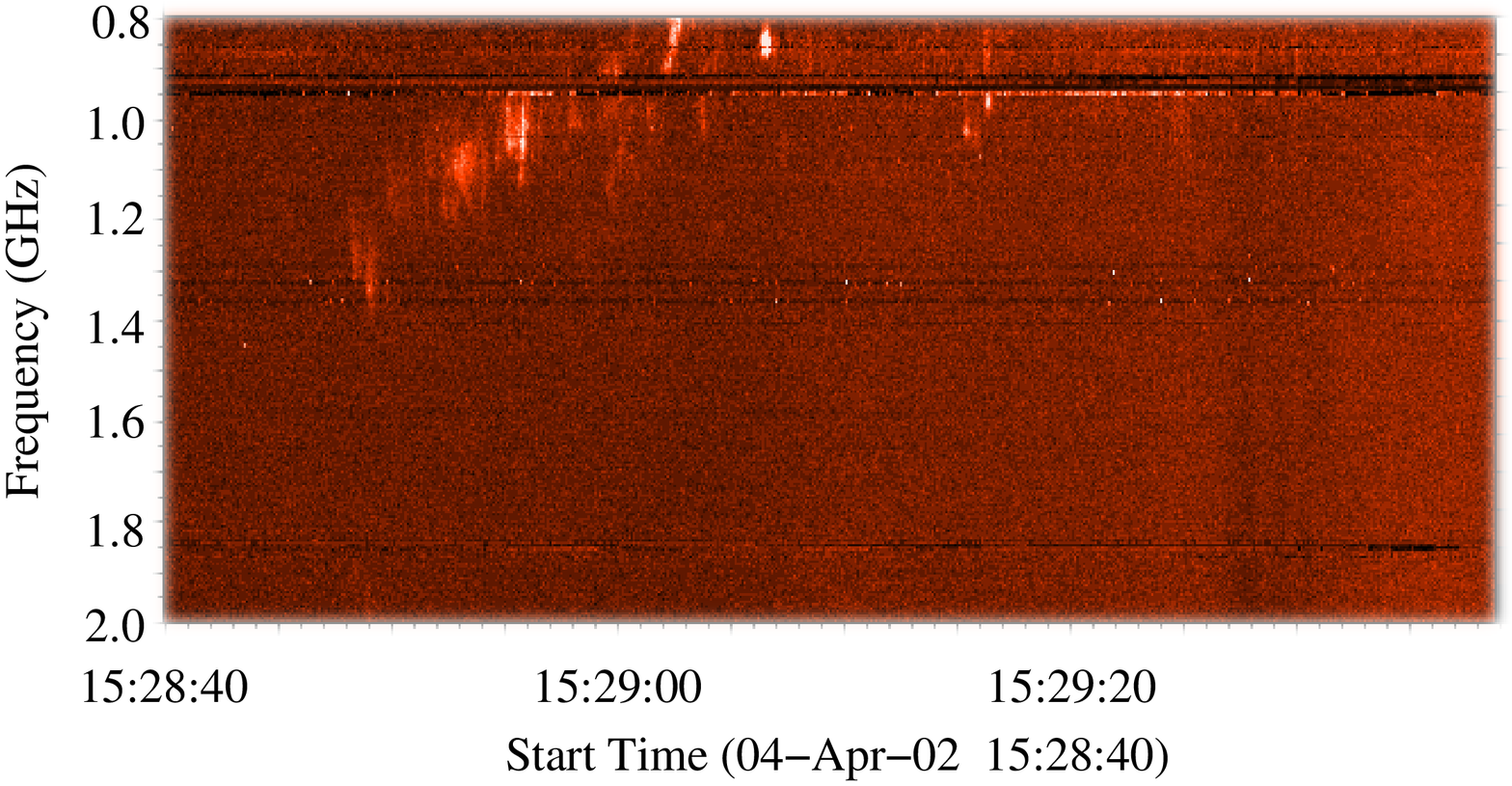}
\caption{Examples of the drifting pulsation structures (DPSs) observed in the
decimetric range during solar flares by the Ond\v{r}ejov
radiospectrograph~\citep{2008SoPh..253...95J}.} \label{Fig2}
\end{figure*}

\section{Introduction}

Solar radio bursts are generally divided into five groups designated as type I,
II, III, IV and V~\citep{1965sra..book.....K}. Most of them show fine
structures, especially type IV bursts: fibers, zebra patterns, spikes and
pulsations~\citep{1979GAM....16...23K}.

In the present paper we study two types of bursts: type I
bursts~\citep{1977sns..book.....E} and special type of the pulsations called
drifting pulsation structures
(DPSs)~\citep{2000A&A...360..715K,2001A&A...369.1104K,2002A&A...395..677K,2004A&A...417..325K,2015ApJ...799..126N}.

Type I radio bursts (noise storms) are quite common phenomenon observed in
metric wavelength range up to about 400 MHz. They last from a few hours to
several days and they are associated with the active region passage over solar
disc \citep{1977sns..book.....E,1995SoPh..160..151K}. They appear in clouds of
short duration and narrow bandwidth bursts (0.1 - 3 s; several MHz),
superimposed on a broadband continuum. The polarization of type I bursts is in
most cases the same as the background continuum polarization. While the
polarization up to 100\% is observed for the noise storms located close to the
solar disc center, at the solar limb their polarization is
lower~\citep{1985srph.book..415K}. The type I bursts are structured, they form
chains of type I bursts (in short type I chains)~\citep{1970A&A.....5..372E}.
These chains preferentially drift towards lower frequencies and their drift is
used for estimations of the magnetic field in their sources
~\citep{1986BAICz..37..115G,1994AuJPh..47..811S,2015SoPh..290..159S}. Sometimes
the type I chains even oscillate in frequencies ("wavy" appearance on dynamic
spectrum)~\citep{1977sns..book.....E}.

Recently,~\citet{2015A&A...576A.136M} have found that noise storms consist of
an extended halo and several compact cores which intensity is changing over a
few seconds. Regions, where storms were originated, were much denser than the
ambient corona and their vertical extent was smaller than estimated from
hydrostatic equilibrium.

Moreover,~\citet{2015ApJ...809...73M} have proposed that persistent magnetic
reconnection along quasi-separatrix layers of the active region is responsible
for the continuous metric noise storm.

Several models of type I bursts were proposed, see the book of
~\citet{1977sns..book.....E} and~\citet{1985srph.book..415K}. Among them the
most promising models are those based on the plasma emission mechanism, e.g.,
the model by ~\citet{1982A&A...105..221S} connecting the type I chains with
weakly super-Alfvenic shocks generated in the front of emerging magnetic flux.

In solar flares, in the decimetric range, pulsations are quite common
\citep{1986SoPh..104...57A,1979GAM....16...23K,1994A&AS..104..145I,2001A&A...375..243J,2007LNP...725..251N,
2004psci.book.....A}. Among them a special type of the pulsations called now
drifting pulsation structures (DPSs) have been recognized and interpreted
\citep{2000A&A...360..715K,2001A&A...369.1104K,2002A&A...395..677K,2002A&A...388..363K,2004A&A...417..325K,
2015ApJ...799..126N}. They are relatively narrowband and drifting mostly
towards lower frequencies. They are usually observed during the impulsive phase
of eruptive flares in the 0.6 - 3 GHz frequency range in connection with the
plasmoid ejection.

Nice example of the plasmoid ejection, observed in soft X-rays during the 5
October 1992 event,  is described in the paper by \citet{1998ApJ...499..934O}.
It is is shown that the plasmoid is a small part of the 3-D loop (Figures 2,
5a, and 10 in \citet{1998ApJ...499..934O}). Further analysis showed that the
plasmoid is a small part of the 3-D current-carrying loop which is embedded in
the current sheet, where the flare magnetic reconnection takes place (see
Figure 2 in \citet{2012ApJ...754...13S}). The magnetic reconnection accelerates
superthermal electrons
\citep{2005PhRvL..94i5001D,2006Natur.443..553D,2006JGRA..11110212P,2008PhPl...15j2105P}
that are then trapped in a denser O-type magnetic structure, which thus becomes
"visible" in the soft X-rays \citep{1998ApJ...499..934O}, EUV
\citep{2012ApJ...745L...6T} or at 17 GHz radio waves
\citep{2010SoPh..266...71K} as the plasmoid.  A limited extent of the plasmoid
along the loop (i.e. trapping of superthermal electrons along the loop) can be
explained by a complexity of magnetic field lines at the plasmoid as shown in
the 3-D kinetic simulation of the magnetic reconnection
\citep{2015ApJ...806..167G} or by the distribution of superthermal electrons
with the high-pitch angles only. Further possibilities of this trapping are
discussed in the paper by \citet{1998ApJ...499..934O}.  The plasmoid in 3-D has
a cylindrical form, which in its 2D models (invariant in the third coordinate)
corresponds to circular magnetic structure having the plasma density greater
than that in the surrounding plasma, see
\citet{2008A&A...477..649B,2008SoPh..253..173B} and the following
Figure~\ref{Fig3}.

The model of DPSs was developed in papers by
\citet{2000A&A...360..715K,2008A&A...477..649B,2011ApJ...737...24B,2010A&A...514A..28K,2011ApJ...733..107K}.
In the 5 October 1992 event the frequency of DPS was found to be close to the
plasma frequency derived from the plasmoid density, see Figure 1 in the paper
by \citet{2008SoPh..253..173B}. Therefore in the DPS model the plasma emission
mechanism is considered. In the flare current sheet, during the magnetic
reconnection plasmoids are generated due to the tearing mode instability. At
X-points of the magnetic reconnection superthermal electrons are accelerated
\citep{2008ApJ...674.1211K}. Then these electrons are trapped in a nearby
plasmoid (O-type magnetic field structure), where they generate plasma waves
that after their conversion to electromagnetic waves produce DPSs (see Figure 9
in \citet{2010A&A...514A..28K}). The narrow bandwidth of DPSs is given by the
limited interval of the plasma densities (plasma frequencies) inside the
plasmoid. In the flare current sheet plasmoids preferentially move upwards in
the solar atmosphere (due to a tension of the surrounding magnetic field lines
\citep{2008A&A...477..649B}), i.e. in the direction, where the electron plasma
density decreases, that is why DPSs mostly drift to low frequencies. The
velocity of plasmoids is in the range from zero to the local Alfv\'en speed.
The acceleration of electrons at X-points of the magnetic reconnection is
quasi-periodic, which causes quasi-periodic pulsations of DPSs. The typical
period of the pulses is about one second. In some cases these pulses have the
frequency drift which is caused by propagation of the superthermal electrons
inside the plasmoid.

In the paper we compare chains of type I bursts and drifting pulsation
structures observed in the impulsive phase of eruptive flares. Based on this
comparison we propose a new model of the chains of type I bursts. The chains of
type I bursts are considered to be radio signatures of processes that heat the
solar corona. Therefore, a correct model of type I bursts can contribute to a
solution  of the problem of the hot corona.

The paper is structured as follows: In Section 2 we compare chains of type I
bursts and DPSs and show their similarities and differences. Then in section 3,
for chains of type I bursts we propose a model similar to that of DPSs and then
we discuss processes which could explain their differences. Conclusions are in
Section 4.

\section{Comparison of chains of type I radio bursts and drifting pulsating structures}

Examples of type I drifting chains and drifting pulsation structures (DPSs) are
shown in Figures~\ref{Fig1} and~\ref{Fig2}. Their typical parameters are
summarized in Table~\ref{tab1}. The parameters of the type I chains were taken
combining the results presented by
\citet{1977sns..book.....E,1979GAM....16...23K,1985srph.book..415K,2015A&A...576A.136M,2015SoPh..290..159S}.
On the other hand, the parameters of DPSs are taken from papers by
\citet{2001A&A...375..243J,2008SoPh..253..173B,2015ApJ...799..126N}. Remark:
For DPSs there is only one polarization measurement (P $\sim$ 30 $\%$),
presented in the paper by \citet{2001A&A...369.1104K}. The brightness
temperature of DPS was calculated for the 5 October 1992 event
\citep{1998ApJ...499..934O,2000A&A...360..715K} assuming that the DPS source
size is equal to the plasmoid size. While for type I chains it is commonly
believed that their polarization is consistent with the O-mode
\citep{1985srph.book..415K}, for DPS it is unclear. Periods of repetition of
type I bursts in chains and pulsations in DPSs are similar, from fractions of
second to several seconds.

\begin{table*}
\caption{Typical parameters of the type I chains and DPSs. T$_{\rm b}$ means
the brightness temperature.} \label{tab1} \centering
\begin{tabular}{cccccccc}
\hline\hline
 & Frequency & Bandwidth & Duration & Drift rates & Polarization & T$_{\rm b}$ & Size  \\
 & range (MHz)     &   (MHz)   &   (s)    &  (MHz s$^{-1}$) & \%  &   (K)   &   (arcsec) \\
 \\
\hline
\\
Type I chain & 30 -- 400 & 5 -- 20 & 10 -- 200 & -1.0 -- +0.3 & up to 100  & $\leq$ 10$^{10}$ & $\sim$ 40 \\
\\
\hline
\\
DPS & 600 -- 3000 & $\sim$ 200 & $\sim$ 60 & -10 -- +5 & $\sim$ 30 & $\sim$ 8 $\times$ 10$^9$ & $\sim$ 10 \\
\\
\hline
\label{tab1}
\end{tabular}
\end{table*}

As shown in Table~\ref{tab1} some parameters of type I chains and DPSs are
comparable (duration and brightness temperature), other parameters differ.
However, type I chains are observed in metric range and DPSs in decimetric
range, i.e. in different altitudes of the solar atmosphere with different
densities, different density gradients and magnetic field strengths. Moreover,
while DPSs are usually observed during the impulsive phase of eruptive
flares~\citep{2015ApJ...799..126N}, type I chains are a part of noise storms
connected with the reconnection activity at the quasi-separatrix layer in
active region~\citep{2015ApJ...809...73M}.

As concerns an appearance of the chains of type I bursts and DPSs
(Figures~\ref{Fig1} and~\ref{Fig2}) they look to be similar. The both types of
bursts preferentially drift towards lower frequencies. In rare cases both
reveal "wavy" appearance, see the chain of type I bursts observed in July 2,
2012 (Figure~\ref{Fig1}, the upper right part) and DPS observed in June 30,
2002 (Figure~\ref{Fig2}, the upper right part).

\section{New model of type I chains}

\begin{figure*}
\centering
\includegraphics[scale=0.5]{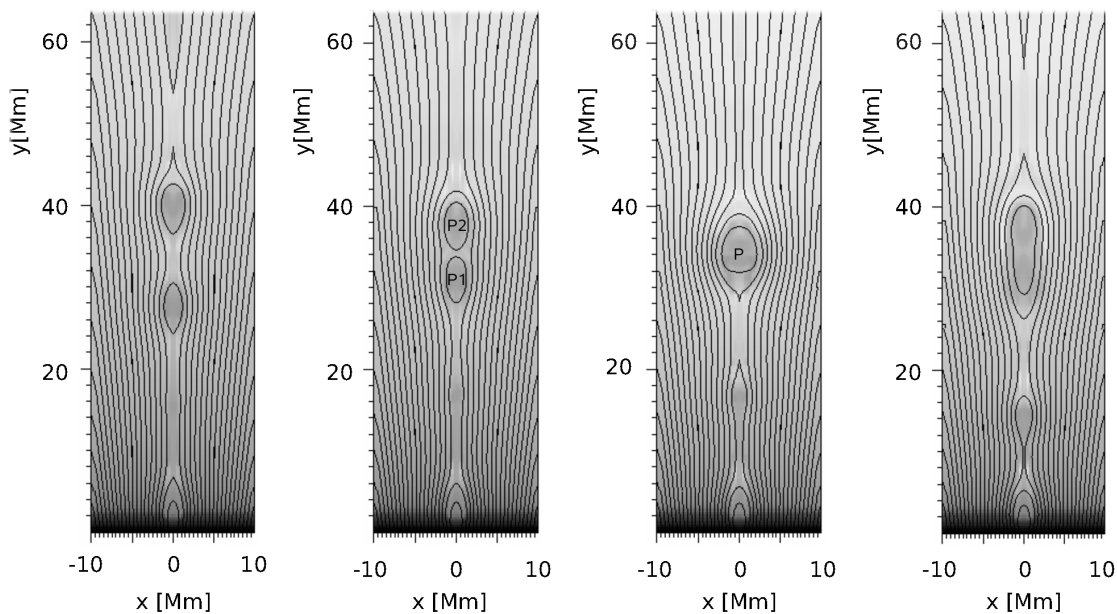}
\caption{Magnetic field lines and densities (grey shades) showing merging of
two plasmoids (P$_1$ and P$_2$) into larger plasmoid P which oscillates. The
panels (from left to right) are shown for t = 80, 90, 100 and 110 s,
respectively.} \label{Fig3}
\end{figure*}

Considering all observational parameters of the both burst types and
differences in conditions in their generation we propose that the type I chain
is produced by very similar processes as DPS; they differ only in plasma
parameters and initial conditions.  Therefore the model of the drifting chain
of type I bursts can be explained as follows:

During the magnetic reconnection in the current sheet formed in the
quasi-separatrix layer of the active region the current-carrying loops are
generated. In the magnetic reconnection between interacting current-carrying
loops there are the X-magnetic points, where electrons are accelerated. These
electrons penetrate into interacting loops at their interaction region. This
region (plasmoid) is simultaneously rapidly heated. For the acceleration
process and penetration of electrons to interacting current-carrying loops, see
Figure 3 in the paper by \cite{2008ApJ...674.1211K}.

As seen in Figures  2, 5a, 10 in the paper by \citet{1998ApJ...499..934O}, the
observed plasmoid in the DPS case is a small part of the loop. We assume that a
similar spatially limited plasmoid is also formed in the case of the drifting
chain of type I bursts. The limited extent of the plasmoid means that the hot
plasma (observed in soft X-rays in the DPS case) is trapped by some processes
in this plasmoid also in direction along its axis. Some processes explaining
this trapping were proposed in the paper by \cite{1998ApJ...499..934O}. But
based on the papers by \cite{2015ApJ...806..167G} and
\cite{2016GeoRL..4310557F} we think that this trapping is mainly due a
complexity of magnetic field lines in the plasmoid. Considering the
observational evidence about the hot plasma confinement in the plasmoid, we
assume that also superthermal electrons (accelerated during the interaction of
the current-carrying loops) are trapped in the plasmoid. (Note that in the
perpendicular direction to the plasmoid axis the hot plasma as well as
superthermal electrons are kept by the magnetic field in the plasmoid.)

These trapped superthermal electrons generate in the plasmoid the Langmuir
(electrostatic) waves, which are then transformed into the electromagnetic
(radio) waves, observed as type I chains. These processes are shown in details
in the paper by \cite{2010A&A...514A..28K}. There, using a 2.5-D
particle-in-cell model, we self-consistently described not only the interaction
of plasmoids, but also how the electrostatic (Langmuir) and electromagnetic
(radio) waves are generated. We found that the distribution of superthermal
electrons, penetrating the plasmoid, is unstable for the Langmuir waves (due to
the bump-on-tail instability) and then these electrostatic waves are
transformed to the electromagnetic (radio) waves, see Figure 9 in this paper.
In agreement with the models of type I bursts based on the plasma wave
theories~\citep{1977sns..book.....E}, we assume that the emission of type I
bursts is dominant on the fundamental frequency and the processes (e.g. the
coalescence of two Langmuir waves) giving the emission on the harmonic
frequency are not effective. The observed polarization of type I bursts is up
to 100 percent. To reach such high polarization, in agreement with
\cite{1963PASJ...15..462T}, and \cite[][page 271]{1980gbs..bookR....M}, we
assume that the type I burst emission is in O-mode generated in the region
where the emission frequency $\omega$ is greater than the plasma frequency
$\omega_{pe}$ and smaller than the cutoff frequency for the X-mode $\omega_x$.
This O-mode (electromagnetic one) is generated from Langmuir waves by
scattering on thermal ions and in such a case the frequency of the resulting
emission is essentially equal to that of Langmuir waves ($\omega_l =
\omega_{pe} (1 + 3 v_{te}^2/v_{\phi}^2)^{1/2}$, where $v_{te}$ is the thermal
plasma velocity and $v_{\phi}$ is the phase velocity). Now using the commonly
accepted assumption that in the type I burst source the electron gyro-frequency
$\omega_{ce}$ is much smaller that the plasma frequency, we have $\omega_x
\approx$ $\omega_{pe}$ + 1/2 $\omega_{ce}$. Then the above mention condition
$\omega \cong \omega_l \leq \omega_x$  gives a lower limit for the phase
velocities of generated Langmuir waves, expressed as $v_{\phi} \geq v_{te} (3
\omega_{pe}/\omega_{ce})^{1/2}$. Because the Langmuir waves with the phase
velocity $v_{\phi}$ are produced by electrons having velocities greater than
$v_{\phi}$, it also gives a lower limit for energies of energetic electrons.

Due to a limited interval of densities inside the plasmoid, the trapped
superthermal electrons generate Langmuir waves in the limited interval of
plasma frequencies and thus the instantaneous bandwidth of type I chain is
limited. Because the plasmoid expands or moves upwards in the solar atmosphere,
plasma densities inside the plasmoid decrease, and therefore most of type I
chains drift towards lower frequencies. Similarly as in DPSs, the acceleration
of superthermal electrons is quasi-periodic which leads to quasi-periodic
repetition of type I bursts (which form the type I chain) with the typical
period of about one second. If accelerated electrons are trapped in several
plasmoids simultaneously then several type I chains are simultaneously
generated. Thus, type I chains can be mutually superimposed in the radio
spectrum (if there are similar densities in the plasmoids) or they are
separated in frequencies in the radio spectrum, if densities inside plasmoids
are different.

As already mentioned, in some cases of type I chains and DPSs there is also
similarity in their "wavy" appearances. Up to now this feature was not
explained in any model of type I chains.

In the DPS model, during the flare magnetic reconnection below the rising
magnetic rope, plasmoids are formed due to the tearing mode
instability~\citep{2000A&A...360..715K,2014RAA....14..753K}. The plasmoids can
merge to larger plasmoids which after this merging (coalescence) process start
to oscillate with the period $P \sim L/v_{\rm A}$, where $L$ is the
characteristic length in the merging process and $v_{\rm A}$ is the local
Alfv\'en speed~\citep{1987ApJ...321.1031T}. Oscillations of the plasmoid
(compression and expansion) periodically change densities inside the plasmoid
and thus periodically change the plasma frequencies and frequencies of DPS.
Just this process was proposed for explaining of quasi-periodic variations of
frequencies ("wavy appearance") of DPSs~\citep{2016CEAB...40...93K}. Here the
same process is proposed for explanation of the "wavy" appearance of type I
chains.

\begin{figure}[h!]
\centering
\includegraphics[width = 0.49\textwidth]{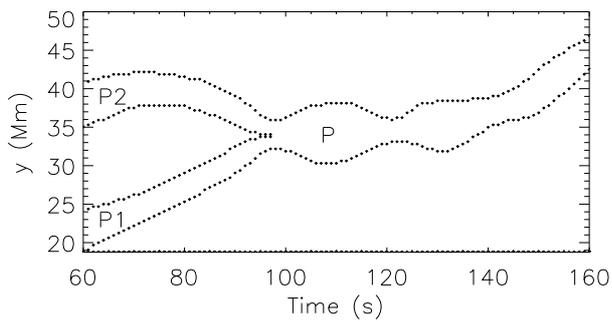}
\caption{Plot showing the merging process of two plasmoids (P$_1$ and P$_2$) along the axis of the vertical and gravitationally
stratified current sheet to one larger and oscillating plasmoid P, compare with Figure~\ref{Fig3}.} \label{Fig4}
\end{figure}

\begin{figure}[h!]
\centering
\includegraphics[width = 0.49\textwidth]{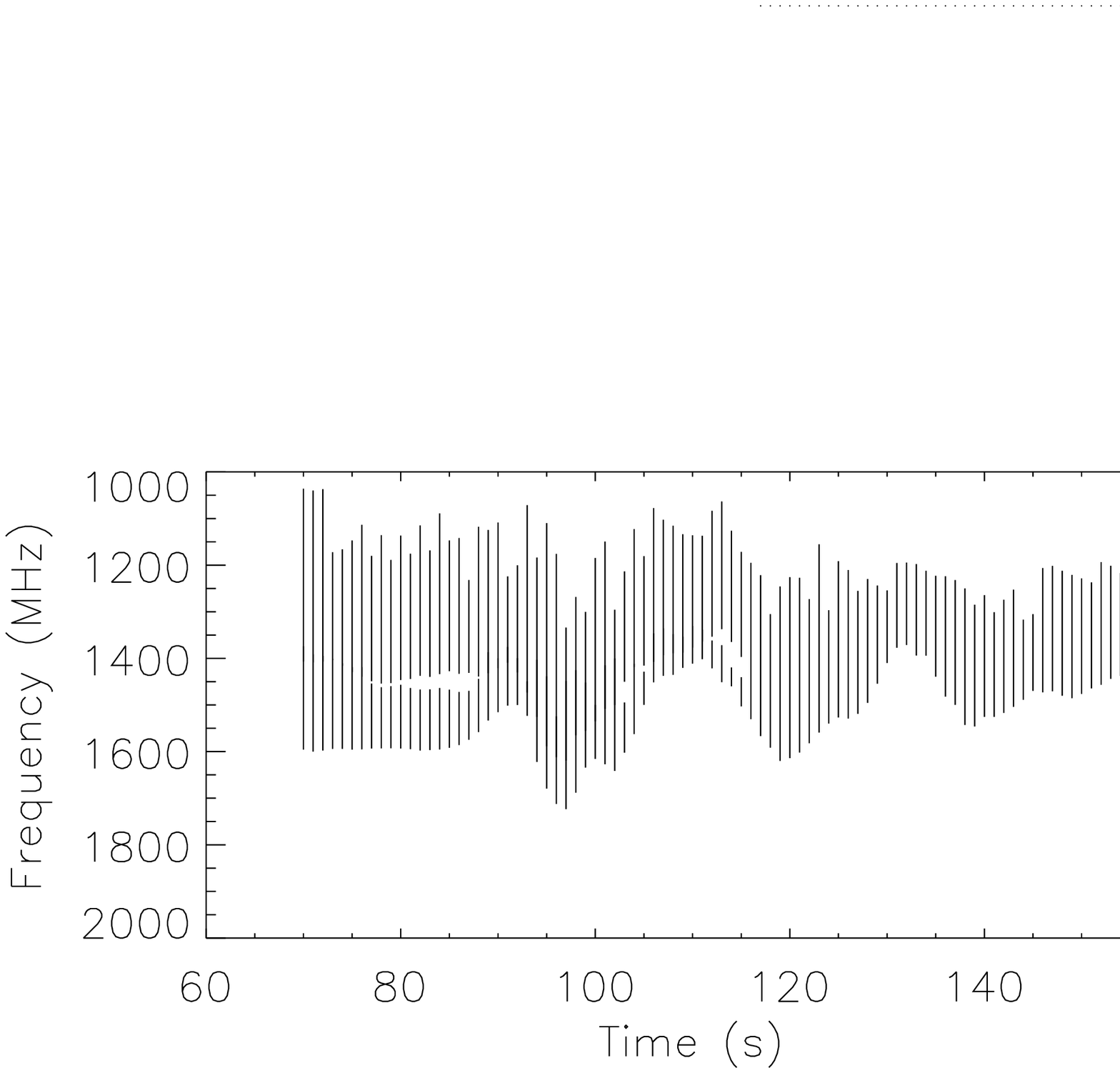}
\caption{The artificial radio spectrum of DPS with the "wavy" appearance corresponding to merging
process of two plasmoids into one larger and oscillating plasmoid as presented in Figures~\ref{Fig3} and~\ref{Fig4}.
Vertical lines mimics pulses in DPS with  the typical period of DPS (about 1 s).} \label{Fig5}
\end{figure}

To illustrate it, we made similar numerical simulations as in the paper
by~\citet{2008A&A...477..649B}, see Figure~\ref{Fig3}, where two plasmoids
(P$_1$ and P$_2$) merge into one larger plasmoid P. After merging the resulting
plasmoid starts to oscillate. Figure~\ref{Fig4} shows time evolution of this
merging and oscillating process expressed by positions of the selected magnetic
field lines (with fixed magnetic vector potential) at the top and bottom of the
plasmoids at the axis of the vertical and gravitationally stratified current
sheet (x = 0 in Figure~\ref{Fig3}). In our case the period of the oscillation
is about 22 s. From computations we know the maximal and minimal densities
inside plasmoids (which are delimited by the selected magnetic field lines) and
then assuming the emission based on the plasma emission mechanism we computed
the artificial radio spectrum shown in Figure~\ref{Fig5}. The vertical lines in
this DPS spectrum mimic pulses with the typical period of about one second.
Namely, the pulses are generated on the kinetic level of the plasma description
and thus their generation cannot be included into the used magnetohydrodynamic
simulation.

As seen in Figure~\ref{Fig5}, the artificial drifting pulsation structure has
the "wavy" appearance. When the oscillating plasmoid is compressed then the
frequency band of the drifting pulsation structure is shifted to higher
frequencies and vice versa. Compare this artificial drifting pulsation
structure with that, observed in June 30, 2002, shown in Figure~\ref{Fig2} (the
upper right part). For details of computations,
see~\citet{2008A&A...477..649B}.

Because type I chains and DPSs with the "wavy" appearance are relatively rare,
it indicates that the full merging of two plasmoids with comparable sizes into
one larger and oscillating plasmoid is also relatively rare.

In previous models the frequency drift of type I chains is connected with the
the Alfv\'en speed at the source of type I chain, e.g., in the model
by~\citet{1982A&A...105..221S} and thus used for the magnetic field
estimations~\citep{2015SoPh..290..159S}. However, in the new model the speed of
the plasmoid can be in the range from zero up to the local Alfv\'en speed,
see~\citet{2008A&A...477..649B}.

Type I chains  are limited from the high-frequency side at about 400 MHz. It is
commonly believed that it is due to the collisional optical depth of the
emission increases with the frequency to the
square~\citep{1993ASSL..184.....B}. On the other hand, the range of type I
chains (below $\sim$ 400 MHz) shows that these chains are generated at upper
parts of the active-region loops located close to the quasi-separatrix layers
of the active region as proposed by~\citet{2015ApJ...809...73M}.

DPSs are observed on higher frequencies than type I chains, at which in the
"quiet" conditions of the solar atmosphere the plasma emission is fully
absorbed. However, during solar flares the atmosphere becomes highly
inhomogeneous and thus transparent for the plasma emission even in the
decimetric range.

If the size of the resulting plasmoid (L), formed during a merging process, is
estimated (e.g. from the instantaneous frequency bandwidth of the chain and
some density model of the solar atmosphere), then the period (P$_W$) of the
chains with the "wavy" appearance can be used for further estimation of the
magnetic field strength $B$ in the chain radio source (the Alfv\'en speed is
$v_{\rm A} = B/\sqrt{\mu_0 \rho}$ $\sim$ $L/P_w $, where $\mu_0$ is the
magnetic permeability of free space and $\rho$ is the plasma density, that can
be determined from the frequency of the chain, see
also~\citet{1987ApJ...321.1031T}).

There is an important difference in processes generating type I chains and
DPSs. While, in DPSs the magnetic reconnection is forced by the positive
feedback between the magnetic reconnection and plasma inflow, given by the
ejection of the whole flare structure upwards, in the processes generating type
I chains this positive feedback in the magnetic reconnection is missing.

These two regimes (without and with the positive feedback) of the magnetic
reconnection together with plasmoids generating type I chains and drifting
pulsation structures can be seen, e.g., in the 2012 July 12
flare~\citep{2014ApJ...784..144D}. Before the flare (at 15:00 -- 16:16 UT)
chains of type I bursts (noise storm) and then during the flare impulsive phase
(at 16:16 UT), when the flare magnetic rope was ejected, the drifting pulsation
structures were observed.

\section{Conclusions}

As shown in previous section some parameters of type I chains and DPSs are
similar (duration, repetition period of type I bursts in the type I chain and
that of pulses in DPS, preference of the frequency drift towards lower
frequencies, brightness temperature and "wavy" appearance in some cases) and
other parameters like frequency range, bandwidth, frequency drift, source size
and polarization are different. However, the differences can be caused by
different conditions where type I chains and DPSs are generated (different
altitude in the solar atmosphere, different densities, different density
gradients and different magnetic field strengths).

Therefore, considering all these similarities and differences we propose a new
model of chains of type I bursts that is very similar to the DPS model.
Although the magnetic reconnection was already proposed for explanation of
noise storms~\citep{2015ApJ...809...73M}, this new model is more specific about
the processes generating the chains of type I bursts, which are a part of noise
storms. The chains of type I bursts are considered to be radio signatures of
processes that heat the solar corona. Therefore, a correct model of these
processes can contribute to a solution of the problem of the hot solar corona.

We show that the chains of type I bursts can be generated by the magnetic
reconnection associated with plasmoids (parts of current-carrying loops). While
a trapping of accelerated superthermal electrons in a single plasmoid leads to
normal type I chain (without the "wavy" appearance), the trapping of
superthermal electrons in an oscillating plasmoid, which can be the result of
merging of two smaller plasmoids, produces the type I chain with the "wavy"
appearance.

Similarly as in DPSs, individual type I bursts (forming the type I chain) are
generated by quasi-periodic acceleration of superthermal electrons and their
plasma emission. The frequency drift of these individual type I bursts can be
caused by propagation of these superthermal electrons inside the plasmoid.

We think that differences of both these types of bursts are also owing to
different regimes of the magnetic reconnection. While in the case of type I
chains the magnetic reconnection and plasmoid interactions are in the
quasi-separatrix layers of the active region in more or less quasi-saturated
regime, in the case of DPSs the magnetic reconnection and plasmoids formation
and their interactions are forced by the upward motion of the flare magnetic
rope.

The new model can explain the "wavy" appearance of some chains of type I bursts
by the merging of two plasmoids into one larger and oscillating one. This
feature was not up to now explained.

We showed that the chains of type I bursts with the "wavy" appearance can be
used for estimation of the magnetic field strength in their sources.
Unfortunately, examples of the chains of type I bursts with the "wavy"
appearance are rare.

DPSs are generated in deeper and denser layers of the solar atmosphere than
chains of type I bursts. In the "quiet" coronal conditions the plasma radio
emission from these deep and dense layers are absorbed. However, during the
impulsive phase of solar flares these deep and dense layers are strongly
disturbed and thus they become transparent also for the DPS emission.

This new model can also explain the finding that the vertical extent of the
noise storm is smaller than estimated from hydrostatic
equilibrium~\citep{2015A&A...576A.136M}. Namely, complicated magnetic field
structure in the region with plasmoids can shorten the density scale-height
similarly as was proposed for microwave type III pair bursts
by~\citet{2016ApJ...819...42T,2016RAA....16e..13T}. Furthermore, this model
explains an enhanced density in the noise storm source comparing to the ambient
corona~\citep{2015A&A...576A.136M}. As was shown, the plasmoids, where the type
I chains are generated, are denser than the surrounding plasma.

In this new model the plasmoid velocity, which is assumed to be connected with
the frequency drift of the type I chain, is in the range from zero to the local
Alfv\'en speed; contrary to previous models, where the velocity of the agent
producing the frequency drift were strictly the Alfv\'en speed. It should be
taken into consideration when the frequency drift of type I chains is used for
magnetic field estimations.

There are still many questions, especially about the plasmoid formation, its
magnetic structure and evolution of the superthermal electrons in real
3-dimensional configuration. For their answers new simulations in extended 3-D
kinetic models are necessary.

\begin{acknowledgements}
The author thanks the referee for constructive comments that improved the
paper. He acknowledges support from Grants 16-13277S and 17-16447S of the Grant
Agency of the Czech Republic.
\end{acknowledgements}

\bibliographystyle{aa}

\end{document}